\documentclass{article}
\usepackage{microtype}
\usepackage{graphicx}
\usepackage{subcaption}
\usepackage{booktabs} 
\usepackage{hyperref}

\usepackage[accepted]{icml2026}
\usepackage{amsmath}
\usepackage{amssymb}
\usepackage{mathtools}
\usepackage{amsthm}
\usepackage{tikz}
\usetikzlibrary{positioning}
\usetikzlibrary{shapes.geometric}
\newcommand{\Tzero}{$T_0$}
\newcommand{\Tone}{$T_1$}
\newcommand{\Ttwo}{$T_2$}
\usepackage[capitalize,noabbrev]{cleveref}
\theoremstyle{plain}

\theoremstyle{definition}

\theoremstyle{remark}

\usepackage[textsize=tiny]{todonotes}
\icmltitlerunning{Position: Early-Stage Quality Assurance in Annotation Pipelines\\Is More Cost-Effective Than Late-Stage Validation}

\begin{document}

\twocolumn[
\icmltitle{Position: Early-Stage Quality Assurance in Annotation Pipelines\\Is More Cost-Effective Than Late-Stage Validation}

\begin{icmlauthorlist}
\icmlauthor{Sunil Kothari}{cair-redmond}
\icmlauthor{Sumukha Sharma Thoppanahalli Chandramouli}{cair-redmond}
\icmlauthor{Naman Khandelwal}{cair-hyd}
\icmlauthor{Parth Kulshreshtha}{cair-chennai}
\icmlauthor{Ashi Jain}{cair-chennai}
\icmlauthor{Kriti Banka}{cair-hyd}
\icmlauthor{Tanuja Chintada}{cair-hyd}
\icmlauthor{Venkata Triveni}{cair-hyd}
\icmlauthor{Gulipalli Praveen Kumar}{cair-hyd}
\icmlauthor{Manish Mehta}{cair-redmond}
\icmlauthor{Tao Liu}{cair-redmond}

\end{icmlauthorlist}

\icmlaffiliation{cair-redmond}{Centific AI Research, Redmond, WA, USA} \icmlaffiliation{cair-hyd}{Centific AI Research, Hyderabad, India} \icmlaffiliation{cair-chennai}{Centific AI Research, Chennai, India}

\icmlcorrespondingauthor{Sunil Kothari}{sunil.kothari@centific.com}
\icmlcorrespondingauthor{Sumukha Sharma Thoppanahalli Chandramouli}{sumukhasharma.t@centific.com}
\icmlcorrespondingauthor{Naman Khandelwal}{naman.khandelwal@centific.com}
\icmlcorrespondingauthor{Parth Kulshreshtha}{parth.kulshreshtha@centific.com}
\icmlcorrespondingauthor{Ashi Jain}{ashi.jain@centific.com}
\icmlcorrespondingauthor{Kriti Banka}{kriti.banka@centific.com}
\icmlcorrespondingauthor{Tanuja Chintada}{tanuja.chintada@centific.com}
\icmlcorrespondingauthor{Venkata Triveni}{nali.v@centific.com}
\icmlcorrespondingauthor{Gulipalli Praveen Kumar}{g.praveenkumar@centific.com}
\icmlcorrespondingauthor{Manish Mehta}{manish.mehta@centific.com}
\icmlcorrespondingauthor{Tao Liu}{tao.l@centific.com}

]
\printAffiliationsAndNotice{\icmlEqualContribution}
\begin{abstract}
This position paper argues that the machine learning community should prioritize early-stage quality assurance in annotation pipelines over the prevailing practice of late-stage validation. Data quality bottlenecks increasingly limit foundation model improvement, yet quality assurance research focuses almost exclusively on validation methods rather than validation timing. \emph{When} validation occurs---not merely \emph{what} validation methods are employed---fundamentally determines both error rates and annotation costs. This temporal neglect is puzzling given the well-established ``shift-left'' principle from software engineering, where empirical studies demonstrate 4--100$\times$ cost multipliers for defects detected in later development stages \citep{boehm1981software, shull2002bugs}. Annotation pipelines, we argue, exhibit analogous dynamics: errors caught before annotation begins cost a fraction of those discovered after review cycles complete.

We propose a taxonomy of three \emph{QA trigger points}---pre-annotation ($T_0$), post-annotation ($T_1$), and post-review ($T_2$)---that decompose annotation workflows into discrete validation opportunities. A parametric error-propagation model formalizes when timing affects final error rates (when stage-specific detection rates differ) versus only economics (when they are equal), making timing a measurable design variable rather than a configuration afterthought. A survey of 47 recent papers reveals that only 4\% report when validation occurs, a striking gap given timing's demonstrated impact in adjacent fields. Without explicit attention to QA timing, the community risks optimizing validation methods while ignoring the structural variable that may matter most.

Acting on this position requires three steps: researchers should report QA timing configurations alongside validation methods; annotation platforms should expose timing as a first-class parameter; and the community should run controlled experiments that measure stage-specific detection rates directly, operationalizing the shift-left principle for annotation pipelines.

\end{abstract}


\section{Introduction}

\textbf{We argue that quality assurance timing---specifically, whether validation occurs before annotation, after annotation, or after review---is a critical design variable that the machine learning community has systematically overlooked, and that early-stage QA is more cost-effective than late-stage validation for most annotation workflows.}

The data-centric AI paradigm has established that training data quality fundamentally constrains model performance \citep{ng2021datacentric, sambasivan2021everyone, whang2023data}. Recent analyses have revealed pervasive label errors across major benchmarks: \citet{northcutt2021pervasive} identified error rates averaging 3.3\% in ImageNet, CIFAR, and other widely-used test sets, while \citet{beyer2020imagenet} found that 6\% of ImageNet validation labels require correction. For video annotation pipelines processing thousands of hours of footage at frame rates of 16--30 fps, even small per-frame error rates compound into substantial quality degradation \citep{dave2020tao, voigtlaender2019mots}.

The research community has responded with increasingly sophisticated validation methods. Confident learning enables statistical detection of label errors without ground truth \citep{northcutt2021confident}. Vision-language models such as GPT-4V and Qwen2-VL provide semantic verification capabilities \citep{wang2024qwen2vl, openai2023gpt4v}. Multi-annotator consensus methods model annotator reliability and aggregate labels accordingly \citep{dawid1979maximum, raykar2010learning, goh2022crowdlab}. Inter-annotator agreement metrics quantify annotation consistency \citep{artstein2008survey, cohen1960coefficient, krippendorff2011computing}.

These advances focus exclusively on \emph{what} to validate and \emph{how} to detect errors. Yet they share a critical blind spot: the question of \emph{when} validation should occur in the annotation pipeline receives virtually no systematic attention. We surveyed 47 articles on CVPR, NeurIPS and ICML annotation quality (2022--2024) and found that only 2 articles (4\%) explicitly reported the stage of the pipeline at which validation was applied.

This omission matters because the same validation method produces different outcomes depending on when it executes. Consider a video annotation pipeline in which machine learning pre-annotation produces bounding boxes with a 15\% error rate. If quality assurance runs only after human review (late-stage), these errors propagate throughout the pipeline before detection, consuming annotator time on corrections that could have been prevented. If quality assurance runs before human annotation begins (early-stage), errors are caught before human effort is invested, potentially reducing both error rates and costs.

Software engineering learned this lesson decades ago. The ``shift-left'' principle, formalized by \citet{boehm1981software} and validated in subsequent empirical studies \citep{shull2002bugs, jones2008economics, mcconnell2004codecomplete}, establishes that defects caught in early development stages cost substantially less to remediate than those caught later. Manufacturing quality management and ML technical debt research encode similar intuitions, which we examine in Section~\ref{sec:evidence}.

We hypothesize that annotation pipelines exhibit dynamics analogous to software development and manufacturing: errors detected early cost less to remediate than errors detected late. This paper does not prove this hypothesis empirically---such proof requires controlled experiments comparing identical validators across pipeline stages, which we have not conducted. Rather, we argue that the hypothesis is sufficiently plausible, and the current neglect of timing sufficiently systematic, that the research community should treat QA timing as a first-class research question.

The remainder of this paper proceeds as follows. Section~\ref{sec:trigger} introduces our taxonomy of QA trigger points and situates it within annotation workflow structure. Section~\ref{sec:evidence} presents evidence that timing is currently invisible in both research and practice. Section~\ref{sec:model} develops a parametric model clarifying when timing affects error rates versus only economics. Section~\ref{sec:taxonomy} presents the full configuration taxonomy. Section~\ref{sec:alternative} engages substantively with alternative views. Section~\ref{sec:action} presents our call to action, Section~\ref{sec:limitation} describes the limitations of this approach and Section~\ref{sec:conclusion} concludes.


\section{QA Trigger Points: A Framework}
\label{sec:trigger}

Before presenting our argument, we must establish precise terminology for discussing when quality assurance occurs in annotation pipelines. We introduce the concept of \emph{QA trigger points}---discrete moments in the annotation workflow where validation can be invoked---and define three canonical trigger points that capture the structure of typical annotation pipelines.

\subsection{Annotation Pipeline Structure}

Modern annotation pipelines for computer vision tasks typically proceed through three phases \citep{roh2019survey, monarch2021humaninloop}. First, \emph{machine learning pre-annotation} generates initial predictions (bounding boxes, segmentation masks, tracking identities) that serve as starting points for human annotators \citep{yao2012interactive, papadopoulos2017extreme}. Second, \emph{human annotation} refines, corrects, or validates these predictions, with annotators adding missing objects, adjusting boundaries, and correcting labels \citep{su2012crowdsourcing, kovashka2016crowdsourcing}. Third, \emph{human review} provides quality control, with reviewers accepting, rejecting, or requesting revisions to submitted annotations \citep{daniel2018quality, vaughan2017making}.

This three-phase structure creates natural boundaries where quality assurance can intervene. We formalize these as trigger points.

\subsection{Trigger Point Definitions}

\begin{figure*}[t]
\centering
\vspace{1em} 
\begin{tikzpicture}[
    node distance=0.6cm,
    stage/.style={rectangle, draw, rounded corners, minimum width=1.8cm, minimum height=0.7cm, font=\scriptsize},
    trigger/.style={diamond, draw, fill=yellow!30, minimum width=0.8cm, minimum height=0.6cm, font=\scriptsize\bfseries},
    arrow/.style={->, thick}
]
\node[stage, fill=blue!20] (ml) {ML Pre-annot.};
\node[trigger, right=0.3cm of ml] (t0) {\Tzero};
\node[stage, fill=green!20, right=0.3cm of t0] (ann) {Annotation};
\node[trigger, right=0.3cm of ann] (t1) {\Tone};
\node[stage, fill=orange!20, right=0.3cm of t1] (rev) {Review};
\node[trigger, right=0.3cm of rev] (t2) {\Ttwo};
\node[stage, fill=gray!20, right=0.3cm of t2] (out) {Output};

\draw[arrow] (ml) -- (t0);
\draw[arrow] (t0) -- (ann);
\draw[arrow] (ann) -- (t1);
\draw[arrow] (t1) -- (rev);
\draw[arrow] (rev) -- (t2);
\draw[arrow] (t2) -- (out);

\node[below=0.2cm of t0, font=\tiny, align=center] {Pre-annot.\\QA};
\node[below=0.2cm of t1, font=\tiny, align=center] {Post-annot.\\QA};
\node[below=0.2cm of t2, font=\tiny, align=center] {Post-review\\QA};

\end{tikzpicture}
\caption{QA trigger points in annotation pipelines. \Tzero{} occurs after ML pre-annotation but before human work. \Tone{} occurs after annotation but before review. \Ttwo{} occurs after review. Each trigger point enables different validation capabilities and incurs different intervention costs.}
\label{fig:pipeline}
\end{figure*}
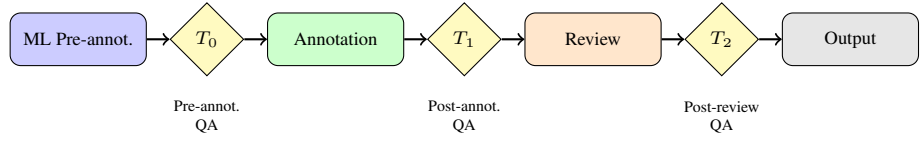

We define three QA trigger points, illustrated in Figure~\ref{fig:pipeline}:

\textbf{Definition 1 (\Tzero: Pre-Annotation Trigger).} The \Tzero{} trigger point occurs after machine learning pre-annotation completes but before human annotators begin work. At \Tzero, validation operates exclusively on machine predictions. QA agents at this stage can assess prediction confidence, detect systematic ML failures, verify coverage against expected object density, and flag anomalous predictions for human attention.

The key characteristic of \Tzero{} is that errors caught here prevent downstream human effort. If a low-confidence ML prediction is flagged at \Tzero{} and routed for re-processing or special handling, annotators never invest time correcting an error that would otherwise propagate through the pipeline.

\textbf{Definition 2 (\Tone: Post-Annotation Trigger).} The \Tone{} trigger point occurs after human annotation submission but before review. At \Tone, validation can compare human annotations against the ML baseline, detect annotation errors (spatial, label, temporal), and---for workflows with multiple annotators---compute inter-annotator agreement (IAA).

The key characteristic of \Tone{} is access to human judgment. Validation at this stage can leverage the comparison between ML predictions and human corrections to identify likely errors. However, human effort has already been invested; errors caught at \Tone{} cannot prevent annotation labor, only review labor.

\textbf{Definition 3 (\Ttwo: Post-Review Trigger).} The \Ttwo{} trigger point occurs after reviewer assessment completes. At \Ttwo, validation can audit reviewer decisions, detect systematic reviewer biases, compare final outputs against gold standards, and generate compliance documentation.

The key characteristic of \Ttwo{} is finality. All human effort (annotation and review) has been invested; errors caught at \Ttwo{} require full rework. However, \Ttwo{} provides the most complete information for validation, including the full provenance chain from ML prediction through annotation to review decision.

\subsection{What Each Trigger Point Enables and Precludes}

The trigger points differ not only in timing but in the validation capabilities they enable (Table~\ref{tab:capabilities}).

\begin{table}[t]
\centering
\caption{Validation capabilities by trigger point. Each stage enables different analyses based on available information.}
\label{tab:capabilities}
\small
\begin{tabular}{@{}lp{5cm}@{}}
\toprule
Trigger & Enabled Capabilities \\
\midrule
\Tzero & ML confidence assessment, coverage validation, systematic failure detection, anomaly flagging \\
\addlinespace
\Tone & Human-ML comparison, annotation error detection, temporal consistency, IAA computation (dual-annotator only), annotator metrics \\
\addlinespace
\Ttwo & Reviewer decision auditing, bias detection, gold standard comparison, compliance documentation \\
\bottomrule
\end{tabular}
\end{table}

Critically, inter-annotator agreement---a primary quality signal in many annotation workflows \citep{artstein2008survey}---is only computable at \Tone{} or later, because it requires completed annotations from multiple annotators. This makes IAA a \emph{timing-dependent} quality signal: workflows that validate only at \Tzero{} cannot leverage agreement as a quality indicator.


\section{Evidence That Timing Is Overlooked}
\label{sec:evidence}

We present three categories of evidence that QA timing is systematically overlooked: a literature survey, an analysis of annotation platforms, and an examination of how adjacent fields treat analogous questions.

\subsection{Literature Survey}

We conducted a structured survey of annotation quality research. Using search terms ``annotation quality,'' ``label quality,'' ``data validation,'' and ``annotation error detection,'' we identified 127 candidate papers from CVPR, NeurIPS, and ICML proceedings (2022--2024). We retained the 47 papers with explicit methodology sections describing validation approaches.

For each paper, we coded: (a) whether validation method characteristics were reported (precision, recall, accuracy); (b) whether computational requirements were reported; (c) whether the pipeline stage at which validation was applied was explicitly stated.

Table~\ref{tab:survey} summarizes our findings. While 100\% of papers reported validation method characteristics and computational requirements, \textbf{only 4.3\% (2/47) explicitly stated when validation was applied}. The two papers that reported timing---both comprehensive surveys by \citet{klie2022annotation} and \citet{klie2024quality}---did so while analyzing annotation practices rather than proposing new methods.\footnote{Complete survey methodology and full paper list available in Supplementary Material.}

\begin{table}[t]
\centering
\caption{Literature survey results: QA timing reporting by venue. Only 2 of 47 surveyed papers explicitly report when validation occurs in the annotation pipeline.}
\label{tab:survey}
\small
\begin{tabular}{@{}lccc@{}}
\toprule
Venue & Papers & Timing Reported & \% \\
\midrule
CVPR/ICCV & 12 & 0 & 0\% \\
NeurIPS & 11 & 0 & 0\% \\
ICML & 8 & 0 & 0\% \\
Other ML venues & 16 & 2 & 12.5\% \\
\midrule
\textbf{Total} & \textbf{47} & \textbf{2} & \textbf{4.3\%} \\
\bottomrule
\end{tabular}
\end{table}

The most comprehensive annotation error detection survey, \citet{klie2022annotation}, reimplemented 18 error detection methods across 9 datasets. Every method focuses on post-annotation detection (\Tone{} or \Ttwo); none evaluate pre-annotation prevention (\Tzero). \citet{klie2024quality}, analyzing quality management practices across 591 NLP dataset papers, found that systematic quality processes ``are only mentioned rarely.''

\subsection{Platform Analysis}

We examined public documentation for six major annotation platforms: Scale AI, Labelbox, CVAT, Label Studio, Amazon SageMaker Ground Truth, and Appen. We assessed whether each platform exposes QA timing as a named, configurable parameter with documented tradeoffs.

\textbf{Finding: No platform explicitly exposes ``QA timing'' as a first-class parameter.} While platforms offer sophisticated multi-stage workflows, timing emerges implicitly from configuration choices rather than being named, documented, or optimized as a design variable.

Labelbox supports up to 10 workflow stages with AutoQA nodes insertable at any stage---but documentation frames this as workflow configuration, not timing optimization. CVAT provides Ground Truth mode (post-annotation comparison) and Honeypot mode (validation items mixed into annotation queues)---the closest approximation to timing exposure, but without guidance on when each approach is optimal. Label Studio supports webhooks for \texttt{TASK\_CREATED} (\Tzero) and \texttt{ANNOTATION\_CREATED} (\Tone), but native \Ttwo{} support requires the Enterprise edition.

\subsection{The Shift-Left Principle}

The systematic neglect of timing in annotation research contrasts sharply with adjacent fields where timing is a central concern. Software engineering's shift-left principle, formalized by \citet{boehm1981software} in his analysis of software economics, establishes that defect remediation costs increase as defects progress through development stages.

Empirical studies support this principle across multiple contexts. \citet{shull2002bugs} found defects cost 4--5$\times$ more to fix in testing than in design. \citet{jones2008economics} documented cost multipliers ranging from 1$\times$ to 100$\times$ depending on defect type and detection stage. \citet{mcconnell2004codecomplete} synthesized industry data showing 10--25$\times$ cost increases for defects escaping to production.

Manufacturing quality management encodes similar principles. Crosby's 1-10-100 rule \citep{crosby1979quality} posits that prevention costs \$1, detection costs \$10, and correction after failure costs \$100. Toyota's production system, widely studied in operations research \citep{liker2004toyota}, emphasizes ``building in quality'' through early-stage defect prevention rather than end-of-line inspection.

\textbf{We hypothesize that annotation pipelines exhibit analogous dynamics.} The conditions for shift-left to apply are: (1) errors can propagate through stages, (2) later detection requires more rework, and (3) detection capabilities exist at early stages. Annotation pipelines plausibly satisfy all three conditions: ML errors propagate to annotation and review; late detection requires discarding completed work; and confidence scores, coverage analysis, and anomaly detection are feasible at \Tzero.

However, we emphasize that this is a \emph{hypothesis}, not a proven fact. The shift-left principle derives from software and manufacturing contexts that differ from annotation in important ways. Empirical validation---comparing identical validators across \Tzero, \Tone, and \Ttwo---is needed to establish whether the principle transfers.


\section{When Does Timing Affect Outcomes?}
\label{sec:model}

To reason precisely about timing effects, we develop a parametric error propagation model that clarifies when timing affects error rates versus only economics.

\subsection{Model Setup}

Consider an annotation pipeline with ML pre-annotation error rate $e_0$ (the fraction of ML predictions that are incorrect). Let $d_{\text{ann}}$ denote the annotator's natural detection rate (the probability an annotator notices and corrects an ML error without QA assistance), and $d_{\text{rev}}$ denote the reviewer's detection rate.

Without any QA intervention, the final error rate is:
\begin{equation}
e_{\text{final}}^{\text{none}} = e_0 \cdot (1 - d_{\text{ann}}) \cdot (1 - d_{\text{rev}})
\end{equation}

This represents errors that escape both annotator and reviewer detection.

\subsection{QA at Different Trigger Points}

Now suppose QA is applied at a single trigger point with detection rate $d_{T_i}$. The final error rate becomes:

\textbf{QA at \Tzero{} only:}
\begin{equation}
e_{\text{final}}^{T_0} = e_0 \cdot (1 - d_{T_0}) \cdot (1 - d_{\text{ann}}) \cdot (1 - d_{\text{rev}})
\end{equation}

\textbf{QA at \Tone{} only:}
\begin{equation}
e_{\text{final}}^{T_1} = e_0 \cdot (1 - d_{\text{ann}}) \cdot (1 - d_{T_1}) \cdot (1 - d_{\text{rev}})
\end{equation}

\textbf{QA at \Ttwo{} only:}
\begin{equation}
e_{\text{final}}^{T_2} = e_0 \cdot (1 - d_{\text{ann}}) \cdot (1 - d_{\text{rev}}) \cdot (1 - d_{T_2})
\end{equation}

\subsection{Key Insight: Timing Effects Are Conditional}

Examining these equations reveals a critical insight: \textbf{if detection rates are equal across stages ($d_{T_0} = d_{T_1} = d_{T_2}$), timing has no effect on final error rate.}

When detection rates are equal, the multiplicative structure ensures identical outcomes regardless of when QA is applied. The difference is only in \emph{where} errors are caught, which affects economics (human effort invested before detection) but not final quality.

Timing affects error rates \textbf{only when detection rates differ across stages}. This can occur because: (a) ML errors may be more systematic and detectable at \Tzero{} than diverse human errors at \Tone; (b) comparing human annotations against ML baseline at \Tone{} may reveal errors invisible to \Tzero{} analysis; or (c) different validation methods may excel at different stages.

\textbf{We do not know which scenario reflects reality.} Determining whether detection rates differ across stages---and in which direction---requires empirical studies that, to our knowledge, have not been conducted.

\subsection{Economic Effects Are Always Present}

Even when detection rates are equal (and thus error rates are identical), timing affects costs. Let $c_{\text{ann}}$ denote annotation cost per task and $c_{\text{rev}}$ denote review cost per task.

With \Tzero{} QA, errors flagged before annotation save $c_{\text{ann}} + c_{\text{rev}}$ per correctly flagged error. With \Ttwo{} QA, errors flagged after review save nothing---the work is already complete.

This economic effect is why shift-left matters even when final quality is unaffected. However, the economic benefit depends on QA \emph{precision}: false positives at \Tzero{} that incorrectly flag correct predictions can \emph{increase} costs by triggering unnecessary rework.


\section{Configuration Taxonomy}
\label{sec:taxonomy}

The three trigger points, combined with workflow options (single vs.\ dual annotator), generate a space of 14 distinct QA configurations.

\subsection{Workflow Options}

\textbf{Option A (Single Annotator + Reviewer):} One annotator produces annotations; a separate reviewer assesses quality. This is cost-efficient but provides no annotator comparison signal.

\textbf{Option B (Dual Annotator + Reviewer):} Two annotators work independently; a reviewer reconciles differences. This enables IAA computation but approximately doubles annotation cost.

\subsection{Configuration Space}

For each workflow option, any non-empty subset of $\{$\Tzero, \Tone, \Ttwo$\}$ can be activated, yielding 7 configurations per option (Table~\ref{tab:taxonomy}).

\begin{table*}[t]
\centering
\caption{QA configuration taxonomy. The 14 configurations represent every non-empty subset of $\{$\Tzero, \Tone, \Ttwo$\}$ across two workflow options. Agent counts, cost multipliers, and quality ratings are illustrative---derived from team experience with video annotation involving bounding boxes and tracking, not empirical measurements. Actual values depend on annotation type, validator complexity, and platform overhead. The ``Best For'' column provides scenario-level guidance; specific deployment choices should account for domain constraints and quality requirements.}
\label{tab:taxonomy}
\small
\begin{tabular}{@{}llcccp{6.5cm}@{}}
\toprule
Config & Triggers & Agents & Cost & Quality & Best For \\
\midrule
\multicolumn{6}{l}{\textit{Option A: Single Annotator + Reviewer}} \\
A-0 & \Tzero & 5 & 1.0$\times$ & Basic & High-volume, low-stakes data; rapid prototyping \\
A-1 & \Tone & 7 & 1.2$\times$ & Good & Standard production pipelines \\
A-2 & \Ttwo & 8 & 1.3$\times$ & Good & Audit-heavy workflows \\
A-0+1 & \Tzero+\Tone & 12 & 2.2$\times$ & High & Mid-stakes datasets; cost-quality balance \\
A-0+2 & \Tzero+\Ttwo & 13 & 2.3$\times$ & Medium & Weak ML with high audit needs \\
A-1+2 & \Tone+\Ttwo & 15 & 2.5$\times$ & High & Compliance-driven workflows \\
A-MAX & \Tzero+\Tone+\Ttwo & 20 & 3.5$\times$ & Maximum & Single-annotator high-quality datasets \\
\midrule
\multicolumn{6}{l}{\textit{Option B: Dual Annotator + Reviewer (+IAA)}} \\
B-0 & \Tzero & 5 & 2.0$\times$ & Good & Research datasets; consensus matters but cost is constrained \\
B-1 & \Tone+IAA & 8 & 2.4$\times$ & High & Standard research benchmarks \\
B-2 & \Ttwo+IAA & 9 & 2.6$\times$ & High & Subjective tasks needing reviewer reconciliation \\
B-0+1 & \Tzero+\Tone+IAA & 13 & 4.4$\times$ & High & High-quality benchmarks with cost discipline \\
B-0+2 & \Tzero+\Ttwo+IAA & 14 & 4.6$\times$ & High & Specialized domains with weak ML \\
B-1+2 & \Tone+\Ttwo+IAA & 17 & 5.0$\times$ & Maximum & Production datasets for high-stakes ML \\
B-MAX & \Tzero+\Tone+\Ttwo+IAA & 22 & 7.0$\times$ & Ultimate & Safety-critical datasets (medical, autonomous driving) \\
\bottomrule
\end{tabular}
\end{table*}

This taxonomy provides vocabulary for discussing QA design. Rather than describing a pipeline as having ``standard quality controls,'' practitioners can specify ``A-1+2 configuration''---immediately communicating that QA runs at \Tone{} and \Ttwo{} with single annotators.


\section{Alternative Views}
\label{sec:alternative}

We engage substantively with four credible counterarguments to our position.

\subsection{``Validation Method Quality Dominates Timing''}

\textbf{The counterargument:} Investing in better detection methods---higher-precision VLMs, improved confident learning algorithms, more sophisticated IAA metrics---provides larger returns than timing optimization. A superior detector deployed at any stage will outperform an inferior detector at the ``optimal'' stage.

\textbf{Our response:} We find this counterargument partially compelling. In regimes where detection rates vary dramatically across methods but minimally across stages, method improvement will dominate timing optimization. Our model confirms this: when $d_{T_0} = d_{T_1} = d_{T_2}$, timing affects only economics, not error rates.

However, we identify three limitations to this counterargument. First, method improvement and timing optimization are not mutually exclusive; the optimal strategy considers both. Second, timing optimization is often cheaper than method improvement---changing configuration requires no new model development. Third, the counterargument assumes detection rates are equal across stages, which is an empirical question, not an established fact.

We do not claim timing dominates method quality. We claim timing deserves study \emph{alongside} method quality, because we currently lack evidence to assess their relative importance.

\subsection{``The Shift-Left Principle May Not Transfer''}

\textbf{The counterargument:} Software engineering and manufacturing findings may not transfer to annotation contexts. Software defects have different characteristics than annotation errors: software bugs can cascade unpredictably through code paths, while annotation errors are typically localized. Manufacturing defects involve physical materials with nonlinear failure modes, while annotation involves human judgment on defined tasks.

\textbf{Our response:} We acknowledge significant uncertainty about whether shift-left transfers. The principle applies when: (1) errors compound through stages, (2) later detection requires more rework, and (3) early detection is feasible. We believe annotation satisfies these conditions, but we have not proven it empirically.

The counterargument strengthens our call for empirical research. If controlled experiments show that timing has minimal effect in annotation contexts---that the shift-left principle does \emph{not} transfer---we will have learned something important about the disanalogy between annotation and other quality-critical processes. Either outcome advances the field.

\subsection{``Instruction Quality Matters More Than QA''}

\textbf{The counterargument:} \citet{radsch2024quality}, analyzing 57,648 images across 924 annotators, found that ``improving labeling instructions yields higher effects than adding QA steps.'' This suggests upstream prevention (better guidelines) dominates downstream detection (any QA), making timing optimization within QA a second-order concern.

\textbf{Our response:} We agree that instruction quality is critical and potentially underweighted in practice. The R\"adsch et al.\ finding is important and should inform annotation pipeline design.

However, their finding addresses \emph{whether} to add QA, not \emph{when} to add it. Given that organizations will deploy QA---for compliance requirements, audit trails, or error detection---the timing question remains relevant. Moreover, \Tzero{} QA can inform instruction improvement by identifying systematic ML failures that annotators encounter, creating a feedback loop between QA and guideline refinement.

The findings are complementary, not contradictory: improve instructions \emph{and} optimize QA timing.

\subsection{``Empirical Evidence Is Insufficient''}

\textbf{The counterargument:} Position papers should be grounded in empirical findings. Advocating for attention to QA timing without empirical evidence that timing matters is premature. The shift-left analogy is speculative; the error propagation model uses assumed parameters; no controlled experiments compare validators across stages.

\textbf{Our response:} This is the strongest counterargument, and we substantially accept it.

We have not proven that timing matters empirically. Our model is theoretical; our parameters are illustrative; our shift-left hypothesis is untested in annotation contexts. A skeptical reader could reasonably conclude that our position is premature.

However, we offer a meta-argument: \emph{timing is currently invisible, and we cannot assess its importance without making it visible}. The field cannot conduct meta-analyses of timing effects when timing is not reported. Our call for reporting costs nothing and enables future empirical assessment.

If the research community begins reporting timing, conducts controlled experiments, and finds that timing effects are negligible, we will have learned something valuable. If timing effects are substantial, we will have enabled optimization that is currently impossible. Either outcome justifies the modest investment of treating timing as a reportable variable.


\section{Call to Action}
\label{sec:action}

We call on four communities to take specific, actionable steps.

\subsection{For Researchers Publishing Validation Methods}

We urge researchers to report QA timing configuration when publishing validation results. Specifically, state whether validation was applied to ML predictions (\Tzero), human annotations before review (\Tone), or post-review outputs (\Ttwo). This requires minimal effort and enables future meta-analysis. Where resources permit, we encourage measuring whether validation methods perform differently at different stages, reporting results even if differences are minimal.

\subsection{For Dataset Creators}

We urge dataset creators to document QA timing alongside methods. Dataset papers should specify not only \emph{what} quality assurance was applied but \emph{when}. For example: ``Quality assurance included automated bounding box validation at \Tone{} and expert review sampling at \Ttwo, with no \Tzero{} validation of ML pre-annotations.'' This is essential for reproducibility and for understanding dataset quality characteristics.

\subsection{For Annotation Platform Developers}

We urge platform developers to expose timing as a named, configurable parameter. Users should be able to specify which validation methods run at \Tzero, \Tone, and \Ttwo, with documented tradeoffs for each choice. Platforms should provide guidance on latency implications, cost implications, and capability implications of timing choices.

\subsection{For the Research Community}

We urge the research community to conduct controlled timing experiments. The highest-value contribution would be experiments that apply identical validators at \Tzero, \Tone, and \Ttwo{} on the same underlying data, with ground truth labels enabling measurement of true detection rates by stage. This would directly test whether the shift-left hypothesis holds for annotation.


\section{Limitations}
\label{sec:limitation}

We acknowledge several limitations. First, our error propagation model is theoretical; the shift-left hypothesis for annotation remains untested. Second, our examples focus on video annotation with bounding boxes and tracking; generalization to other modalities requires validation. Third, agent counts, cost multipliers, and quality ratings in Table~\ref{tab:taxonomy} are illustrative, not empirically derived. Fourth, our platform analysis is based on public documentation; enterprise features may provide timing configurability. Fifth, our survey may miss papers that mention timing incidentally; the 4\% figure reflects explicit reporting.


\section{Conclusion}
\label{sec:conclusion}

We have argued that quality assurance timing in annotation pipelines---the question of whether validation occurs at \Tzero{} (pre-annotation), \Tone{} (post-annotation), or \Ttwo{} (post-review)---deserves systematic study as a first-class research question.

Our parametric model clarifies when timing affects error rates (when detection rates differ across stages) versus only economics (when detection rates are equal). The shift-left analogy provides theoretical grounding; empirical validation in annotation contexts remains future work.

The evidence suggests timing is currently invisible: surveyed papers rarely report when validation occurs, and no major platform exposes timing as a named parameter. This invisibility prevents the field from assessing timing's importance.

Our call to action is deliberately modest. We ask researchers to report timing when publishing validation results. We ask platforms to expose timing as a configurable parameter. We ask the community to conduct controlled experiments that would operationalize the shift-left principle in annotation contexts.

If these efforts reveal that timing effects are minimal, we will have learned that annotation differs from software and manufacturing in important ways. If timing effects are substantial, we will have enabled optimization currently impossible. Either outcome advances the field beyond the current state where timing is simply ignored.


\bibliographystyle{plainnat}

\begin{thebibliography}{99}

\bibitem[Artstein \& Poesio(2008)]{artstein2008survey}
Artstein, R. and Poesio, M. (2008).
\newblock Survey article: Inter-coder agreement for computational linguistics.
\newblock \emph{Computational Linguistics}, 34(4):555--596.

\bibitem[Beyer et al.(2020)]{beyer2020imagenet}
Beyer, L., H\'enaff, O.~J., Kolesnikov, A., Zhai, X., and van den Oord, A. (2020).
\newblock Are we done with {ImageNet}?
\newblock \emph{arXiv preprint arXiv:2006.07159}.

\bibitem[Boehm(1981)]{boehm1981software}
Boehm, B.~W. (1981).
\newblock \emph{Software Engineering Economics}.
\newblock Prentice-Hall.

\bibitem[Cohen(1960)]{cohen1960coefficient}
Cohen, J. (1960).
\newblock A coefficient of agreement for nominal scales.
\newblock \emph{Educational and Psychological Measurement}, 20(1):37--46.

\bibitem[Crosby(1979)]{crosby1979quality}
Crosby, P.~B. (1979).
\newblock \emph{Quality Is Free: The Art of Making Quality Certain}.
\newblock McGraw-Hill.

\bibitem[Daniel et al.(2018)]{daniel2018quality}
Daniel, F., Kucherbaev, P., Cappiello, C., Benatallah, B., and Allahbakhsh, M. (2018).
\newblock Quality control in crowdsourcing: A survey.
\newblock \emph{ACM Computing Surveys}, 51(1):1--40.

\bibitem[Dave et al.(2020)]{dave2020tao}
Dave, A., Khurana, T., Tokmakov, P., Schmid, C., and Ramanan, D. (2020).
\newblock {TAO}: A large-scale benchmark for tracking any object.
\newblock In \emph{ECCV}.

\bibitem[Dawid \& Skene(1979)]{dawid1979maximum}
Dawid, A.~P. and Skene, A.~M. (1979).
\newblock Maximum likelihood estimation of observer error-rates.
\newblock \emph{J.~Royal Statistical Society C}, 28(1):20--28.

\bibitem[Goh et al.(2022)]{goh2022crowdlab}
Goh, H.~W., Tkachenko, U., and Mueller, J. (2022).
\newblock {CROWDLAB}: Supervised learning for multi-annotator consensus.
\newblock In \emph{NeurIPS Human in the Loop Learning Workshop, 2022}.

\bibitem[Jones(2008)]{jones2008economics}
Jones, C. (2008).
\newblock \emph{Applied Software Measurement}.
\newblock McGraw-Hill, 3rd edition.

\bibitem[Klie et al.(2023)]{klie2022annotation}
Klie, J.-C., Webber, B., and Gurevych, I. (2023).
\newblock Annotation error detection: Analyzing past and present.
\newblock \emph{Computational Linguistics}, 49:157--198.

\bibitem[Klie et al.(2024)]{klie2024quality}
Klie, J.-C., Eckart de Castilho, R., and Gurevych, I. (2024).
\newblock Analyzing dataset annotation quality management.
\newblock \emph{Computational Linguistics}, 50(3):817--866.

\bibitem[Kovashka et al.(2016)]{kovashka2016crowdsourcing}
Kovashka, A., Russakovsky, O., Fei-Fei, L., and Grauman, K. (2016).
\newblock Crowdsourcing in computer vision.
\newblock \emph{Found.~Trends Comput.~Graph.~Vis.}, 10(3):177--243.

\bibitem[Krippendorff(2011)]{krippendorff2011computing}
Krippendorff, K. (2011).
\newblock Computing {Krippendorff's} alpha-reliability.
\newblock Technical report, University of Pennsylvania.

\bibitem[Liker(2004)]{liker2004toyota}
Liker, J.~K. (2004).
\newblock \emph{The Toyota Way}.
\newblock McGraw-Hill.

\bibitem[McConnell(2004)]{mcconnell2004codecomplete}
McConnell, S. (2004).
\newblock \emph{Code Complete}.
\newblock Microsoft Press, 2nd edition.

\bibitem[Monarch(2021)]{monarch2021humaninloop}
Monarch, R.~M. (2021).
\newblock \emph{Human-in-the-Loop Machine Learning}.
\newblock Manning Publications.

\bibitem[Ng(2021)]{ng2021datacentric}
Ng, A. (2021).
\newblock {MLOps}: From model-centric to data-centric {AI}.
\newblock DeepLearning.AI.

\bibitem[Northcutt et al.(2021a)]{northcutt2021confident}
Northcutt, C.~G., Jiang, L., and Chuang, I.~L. (2021a).
\newblock Confident learning: Estimating uncertainty in labels.
\newblock \emph{JAIR}, 70:1373--1411.

\bibitem[Northcutt et al.(2021b)]{northcutt2021pervasive}
Northcutt, C.~G., Athalye, A., and Mueller, J. (2021b).
\newblock Pervasive label errors in test sets.
\newblock In \emph{NeurIPS Datasets \& Benchmarks}.

\bibitem[OpenAI(2023)]{openai2023gpt4v}
OpenAI (2023).
\newblock {GPT-4V(ision)} system card.
\newblock Technical report.

\bibitem[Papadopoulos et al.(2017)]{papadopoulos2017extreme}
Papadopoulos, D.~P., Uijlings, J.~R., Keller, F., and Ferrari, V. (2017).
\newblock Extreme clicking for efficient object annotation.
\newblock In \emph{ICCV}.



\bibitem[R\"adsch et al.(2024)]{radsch2024quality}
R\"adsch, T., Reinke, A., Weru, V., Tizabi, M.~D., Heller, N., Isensee, F., Kopp-Schneider, A., and Maier-Hein, L. (2024).
\newblock Quality assured: Rethinking annotation strategies in imaging {AI}.
\newblock In \emph{ECCV}, pages 52--69.



\bibitem[Raykar et al.(2010)]{raykar2010learning}
Raykar, V.~C. et al. (2010).
\newblock Learning from crowds.
\newblock \emph{JMLR}, 11:1297--1322.

\bibitem[Roh et al.(2019)]{roh2019survey}
Roh, Y., Heo, G., and Whang, S.~E. (2019).
\newblock A survey on data collection for machine learning.
\newblock \emph{IEEE TKDE}, 33(4):1328--1347.

\bibitem[Sambasivan et al.(2021)]{sambasivan2021everyone}
Sambasivan, N. et al. (2021).
\newblock ``Everyone wants to do the model work, not the data work.''
\newblock In \emph{CHI}.

\bibitem[Sculley et al.(2015)]{sculley2015hidden}
Sculley, D. et al. (2015).
\newblock Hidden technical debt in ML systems.
\newblock In \emph{NIPS'15}.

\bibitem[Shull et al.(2002)]{shull2002bugs}
Shull, F. et al. (2002).
\newblock What we have learned about fighting defects.
\newblock In \emph{IEEE International Symposium on Software Metrics}.

\bibitem[Su et al.(2012)]{su2012crowdsourcing}
Su, H., Deng, J., and Fei-Fei, L. (2012).
\newblock Crowdsourcing annotations for visual object detection.
\newblock In \emph{AAAI Workshop}.

\bibitem[Vaughan(2017)]{vaughan2017making}
Vaughan, J.~W. (2017).
\newblock Making better use of the crowd.
\newblock \emph{JMLR}, 18(193):1--46.

\bibitem[Voigtlaender et al.(2019)]{voigtlaender2019mots}
Voigtlaender, P. et al. (2019).
\newblock {MOTS}: Multi-object tracking and segmentation.
\newblock In \emph{CVPR}.

\bibitem[Wang et al.(2024)]{wang2024qwen2vl}
Wang, P. et al. (2024).
\newblock {Qwen2-VL}: Vision-language model perception.
\newblock \emph{arXiv:2409.12191[cs.CV]}.

\bibitem[Whang et al.(2023)]{whang2023data}
Whang, S.~E., Roh, Y., Song, H., and Lee, J.-G. (2023).
\newblock Data collection and quality challenges in deep learning.
\newblock \emph{VLDB Journal}, 32:791--813.

\bibitem[Yao et al.(2012)]{yao2012interactive}
Yao, A., Gall, J., Leistner, C., and Van Gool, L. (2012).
\newblock Interactive object detection.
\newblock In \emph{CVPR}.

\end{thebibliography}

\end{document}